\documentclass[pre,aps,showpacs,12pt]{revtex4}
\usepackage{epsfig}
\begin{document}
\title{Plasticity of Ferromagnets near the Curie Point}
\author{M. Molotskii}
\affiliation{Wolfson Applied Materials Research Centre.\\ Tel Aviv
University, Tel Aviv 69978, Israel} \author{and V.
Fleurov\footnote{Email: fleurov@post.tau.ac.il}}
\affiliation{School of Physics and Astronomy, Beverly and Raymond
Sackler Faculty of Exact Sciences.\\ Tel Aviv University, Tel Aviv
69978, Israel}
\date{\today}

\begin{abstract}
We propose an explanation of the anomalous growth of plasticity in
ferromagnets near the Curie point. We demonstrate that this effect
is caused by spin-dependent detachment of dislocations from
obstacles under an influence of the internal magnetic field.
Magnetization fluctuations grow in the vicinity of the Curie
point, yield an increase of the detachment probability and, hence,
an increase of the plasticity. We apply this model for a
description of the temperature behaviour of the critical stress in
nickel and of the microhardness of gadolinium. An external
magnetic field suppresses the magnetization fluctuations and,
hence may suppress the above singularities.
\end{abstract}
\maketitle
\section{Introduction}

Various experimental data indicate a strong influence of the
magnetic structure of ferromagnets on their plastic properties.
Transition to a magnetically ordered state is accompanied by a
strong change of plasticity (Zackay and Hazlett, 1953,
Nabutovskaya, 1969,1971, Wolfenden, 1978, Bolling and Richman,
1969, Echigoya etal, 1973, Echigoya and Hayashi, 1979, Flor etal,
1980, Retat etal, 1985, Retat, 1987, Maksimova and Maiboroda,
1992, Gulayev and Svistunova, 1996). Measurements with small
temperature intervals revealed an increase of plasticity in a
close vicinity of the Curie point (Zackay and Hazlett, 1953,
Nabutovskaya, 1969,1971, Wolfenden, 1978). This fact by itself is
not so astonishing since many properties of ferromagnets change
near the Curie point (Vonsovskii, 1974), however, the giant value
of the observed effect is really surprising. According to
Nabutovskaya (1969,1971) the microhardness of gadolinium changes
near $T_C$ by a factor of two, whereas its elastic constants vary
by a percent or so. For example, the Young modulus of gadolinium
decreases in the Curie point only by 1.6\% (Spichkin etal, 1999).

Recently we considered a mechanism of the influence of a magnetic
field on plasticity of some nonmagnetic crystals (Molotskii and
Fleurov, 1997, Molotskii, 2000, and references therein). The
magnetic field induces transitions (singlet to triplet) between
different spin states of the radical pairs formed by dangling
bonds of a dislocation cores and paramagnetic obstacles in the
course of the pair formation. The state of the pair may be either
bonding or antibonding, depending on its spin configuration. The
magnetic field can influence the relative occupations of these
states and lead to an increase of probability of the dislocation
detachment from paramagnetic obstacles. The crystal plasticity, as
a result, increases.

This approach appeared to be very successful and allowed for an
explanation of a large number of plasticity related phenomena in
nonmagnetic crystals. We propose here to apply the same approach
when discussing plasticity of ferromagnets. In this case, the role
of the external magnetic field can be played by the local magnetic
induction created by the spontaneously magnetised surrounding
atoms. Recently, we demonstrated that this internal magnetic field
may be important for the Invar hardening, leading to an increase
of the critical resolved shear stress of Invar alloys with a
lowering temperature (Molotskii and Fleurov, 2001). We expect to
observe a temperature variation of various plasticity related
characteristics of the crystal connected with the variation of the
spontaneous magnetization with the temperature. When approaching
the Curie point the role of the magnetization fluctuations may
become more pronounced. The typical time of relaxation of the
magnetization fluctuations is usually much larger than the
characteristic times of formation or rupture of a dislocation -
obstacle bond. Therefore we should consider this process under the
influence of slowly varying magnetic inductions due to fluctuating
magnetization. We will show below that this may result in a
strongly increasing plasticity in the vicinity of the Curie point.

It is worth mentioning here that the theory, we develop in this
paper, will not take into account an interaction of dislocations
with domain walls, which may serve as efficient pinning centres
for dislocation (Seeger etal, 1964). This assumption is justified
as long as the typical domain size  essentially exceeds the
dislocation path length. According to the experiments of Zackay
and Hazlett (1953), and of Nabutovskaya (1969, 1971) this path is
limited by the average distance, $L_f \sim 1/\sqrt{\rho_f}$,
between the forest dislocations. The density of the forest
dislocations in those experiments was $\rho_f \sim 10^9$cm$^{-2}$,
hence the dislocation path length could not exceed $L_f \sim
3\times 10^{-5}$cm, which was two to three orders of magnitude
smaller than the typical domain sizes.

\section{Influence of Internal Magnetic Induction on Plasticity
of Magnets}

Internal magnetic field of a ferromagnet due to spontaneous
magnetization is rather large ($\sim 1$T) and is capable of
influencing plastic properties of crystals. Such an important
characteristic of crystal plasticity as the dislocation path
length increases with the magnetic induction $B$ as (Molotskii,
2000)
\begin{equation}\label{fpl}
  L(B) = L_0(1 + \frac{B^2}{B^2_0})
\end{equation}
where $B_0$ is a constant characterizing the dislocation --
obstacle bond. Usually its value lies in the range from 0.2 to 1T.
In the case of nonmagnetic crystals $B$ is the value of the
magnetic induction created by an external source. In the case of a
ferromagnet we do not need an external magnetic field since a
strong enough internal magnetic induction is always present. This
internal magnetic induction decreases when temperature approaches
the Curie point $T_C$ and starts strongly fluctuating. One may
expect that in the critical region, at $T \rightarrow T_C$, these
strong fluctuations of the internal magnetic induction will
strongly influence plastic properties of the ferromagnet.

The magnetic field influences kinetics of the formation of the
dislocation -- obstacle bonds. The characteristic times of these
processes are typically on the order of $10^{-7}$s (Molotskii and
Fleurov, 1997), which is shorter than the typical times of the
large scale magnetization fluctuations determining the local
internal magnetic induction $B_{int}$ at each bond (see, e.g., Ma,
1976). This allows one to consider the induction $B_{int}$ at each
moment of time as created by a particular space distribution of
the magnetization fluctuations in the ferromagnet.

For example, we estimate here the average dislocation free path
length. According to equation (\ref{fpl}) it is determined by the
mean square of the internal magnetic induction,
\begin{equation}\label{aver1}
\langle B_{int}^2 \rangle = {\overline{B}}_{int}^2 + \langle\Delta
B^2 \rangle.
\end{equation}
Here $\overline{B}_{int}$ is average value of the internal
magnetic induction, whereas $\Delta B = B_{int} -
\overline{B}_{int}$.

Fluctuations of the local internal magnetic induction acting on a
particular dislocation -- obstacle bond can be described using the
theory of the second order phase transitions (see, e.g., Landau
and Lifshitz, 1980). The local internal magnetic indiction of a
homogeneously magnetised crystal can be estimated with the help of
the Lorentz formula
\begin{equation}\label{lorentz}
  {\bf B} = \frac{4\pi}{3}{\bf M}.
\end{equation}
The probability that the magnetization $M$ deviates by $\Delta M$
from its average value $\overline{M}$ in the volume $V$ around the
bond is
\begin{equation}\label{fluct1}
w(\Delta M, V) = \frac{1}{2}\frac{{V^*}^{1/2}}{(2\pi k_B T\chi
)^{3/2}}\left\{
\begin{array}{cc}
\exp \left \{\displaystyle - \frac{{\Delta\bf M}^2V}{2k_BT_C
\chi}\right\}, & V > V^*
\\ 0, & V < V^*
\end{array}\right.
\end{equation}
where $\chi$ is the magnetic susceptibility, $k_B$ is Boltzmann
constant. The probability density function (\ref{fluct1}) is
truncated at volumes smaller than certain characteristic volume
$V^*$ which will later serve as a fitting parameter. The general
approach to the order parameter fluctuations, applied here,
considers the ferromagnet as continuus medium. Hence, it holds at
distances which are larger than typical interatomic distances.
Therefore one may expect the typical scale of the excluded volume
$V^*$ not to exceed essentially several lattice spacings. Similar
approach is implied by equation (\ref{lorentz}), which is obtained
by considering a small empty volume surrounded by a magnetised
medium.

The mean square fluctuations of the magnetization can be then
calculated as
\begin{equation}\label{fluct2}
\langle {\Delta {\bf M}}^2\rangle = \int_{V^*}^\infty d^3 \Delta M
(\Delta {\bf M})^2 w(\Delta M, V) = \frac{k_BT_C\chi}{V^*}
\end{equation}
Now using equations (\ref{aver1}) and (\ref{fluct2}) the mean
square local magnetic field acting on the dislocation -- obstacle
bond can be obtained using the following equations
\begin{equation}\label{aver2}
\langle{{\bf B}_{int}}~^2\rangle = \overline{B}_{int}~^2 + \langle
\Delta B^2 \rangle = \displaystyle\frac{16\pi^2}{9} \overline{{\bf
M}}~^2 + \displaystyle \frac{16\pi^2}{9} \frac{k_BT_C\chi}{V^*} .
\end{equation}

An inhomogeneous spatial distribution of the magnetization, caused
by fluctuations, results in a demagnetization field, which should
have been accounted for in equation (\ref{lorentz}). The
demagnetization field is the strongest for the spherical
fluctuations and is negligible for a fluctuation having the form
of a plate. The exact account of the demagnetization is a very
tedious task but finally leads to a numerical factor, smaller than
one, in the second term in equation (\ref{aver2}). This factor is
absorbed in the volume $V^*$, which anyhow serves as a fitting
parameter.

The temperature dependence of the mean square local magnetic field
in the vicinity of the Curie point can be readily calculated. For
this we may use the temperature dependence of the magnetic
susceptibility which has the form
\begin{equation}\label{suscept}
\chi(T)=\left\{\begin{array}{cc} \displaystyle
\frac{C}{2T_C}\left(\frac{T_C}{T_C-T}\right)^\gamma, & \mbox{at}\
\ \
T<T_C \\
& \\
 \displaystyle \frac{C}{T_C}\left(\frac{T_C}{T -
T_C}\right)^\gamma & \mbox{at}\ \ \ T>T_C.
\end{array} \right.
\end{equation}
Here $\gamma$ is a critical index,
\begin{equation}\label{curie}
  C = \frac{np^2\mu_B^2}{3k_B}
\end{equation}
is the Curie constant, in which $n$ is the particle density, $p$
is the effective number of the Bohr magnetons $\mu_B$ per atom.

The first term in equation (\ref{aver2}) contains the average
spontaneous magnetization $\overline{M}(T)$ whose temperature
dependence is
\begin{equation}\label{aver3}
\overline{M}(T) = \left\{
\begin{array}{cc}
\displaystyle M_0\left(1- \frac{T}{T_C}\right)^\beta & \mbox{at} \
\ \ T < T_C, \\ & \\ 0 & \mbox{at} \ \ \ T > T_C.
\end{array}
\right.
\end{equation}
Here $M_0$ is the limiting value of the magnetization at low
temperature, $\beta$ is the corresponding critical index

\section{Temperature dependence of the critical stress: nickel as a test
case} \label{Ni}

An influence of the magnetization on plasticity was first
discovered half a century ago by Zackay and Hazlett (1953), who
studied the temperature dependence of the critical stress
$\sigma_c$ for nickel. They have found that the dependence of
$\sigma_c(T)$ passes a minimum near the Curie point $T_C$. Samples
with 99.95\% nickel content were used. There was also a small
amount of iron (0.03\%) and magnesium (0.02\%) atoms. The
paramagnetic iron atoms are efficient obstacles for dislocations
in nickel, and they are most probably responsible for the observed
sensitivity of the plasticity of nickel to the magnetic field.

The temperature dependence of the critical stress in a ferromagnet
can be found, accounting for the part played by the internal
magnetic fields. Dislocations can be bound to paramagnetic
obstacles with the binding energy $W_M$. Then the critical stress
is (Friedel, 1964)
\begin{equation}\label{crst1}
\sigma_c (0) = \frac{|W_M|}{b^2 l(B)} \left[1 - \left(
\frac{T}{T_0} \right)^{\frac{2}{3}}\right]^{\frac{3}{2}}
\end{equation}
where $b$ is the value of the dislocation Burgers vector, and $l$
is the average distance between the obstacles. The temperature
dependent factor in equation (\ref{crst1}) accounts for the
thermal activation processes (Haasen, 1983). The parameter $T_0$
is proportional to the dislocation - obstacle binding energy.

It was shown in our paper (Molotskii and Fleurov, 1997) that the
average length of the dislocation free segment depends on the
magnetic field similarly to (\ref{fpl}),
\begin{equation}\label{seg1}
l(B) = l_0 \left(1 + \frac{B^2}{B^2_0}\right).
\end{equation}
Considering a ferromagnet we should introduce the fluctuating
internal local magnetic induction ${\bf B}_{int} = \overline{{\bf
B}}_{int} + \Delta{\bf B}$ which acts on each particular
dislocation -- obstacle bond. It means that we have to average
equation (\ref{seg1}) over various values and orientations of the
vector $\Delta{\bf B}$. This averaging for the case of an easy
axis ferromagnet is carried out in Appendix), so one gets
\begin{equation}\label{seg2}
\frac{\sigma_c(T)}{\sigma_{c0}} = \overline{f}(\overline{B}, \langle
\Delta B^2\rangle) \left[1 - \left( \frac{T}{T_0}
\right)^{\frac{2}{3}}\right]^{\frac{3}{2}}.
\end{equation}
The critical stress appears now to be a function of the average
spontaneous magnetization and of its fluctuations. The temperature
dependence of both these quantities is determined by equations
(\ref{aver2}), (\ref{suscept}), and (\ref{aver3}). As a result we now
have an equation for the temperature dependence of the critical stress
in the vicinity of the Curie temperature.

Figure 1 presents a comparison of the temperature dependence of
the critical stress in Ni calculated by means of equation
(\ref{seg2}) with the available experimental data.
\begin{figure}[htb]
{\includegraphics[width=16cm]{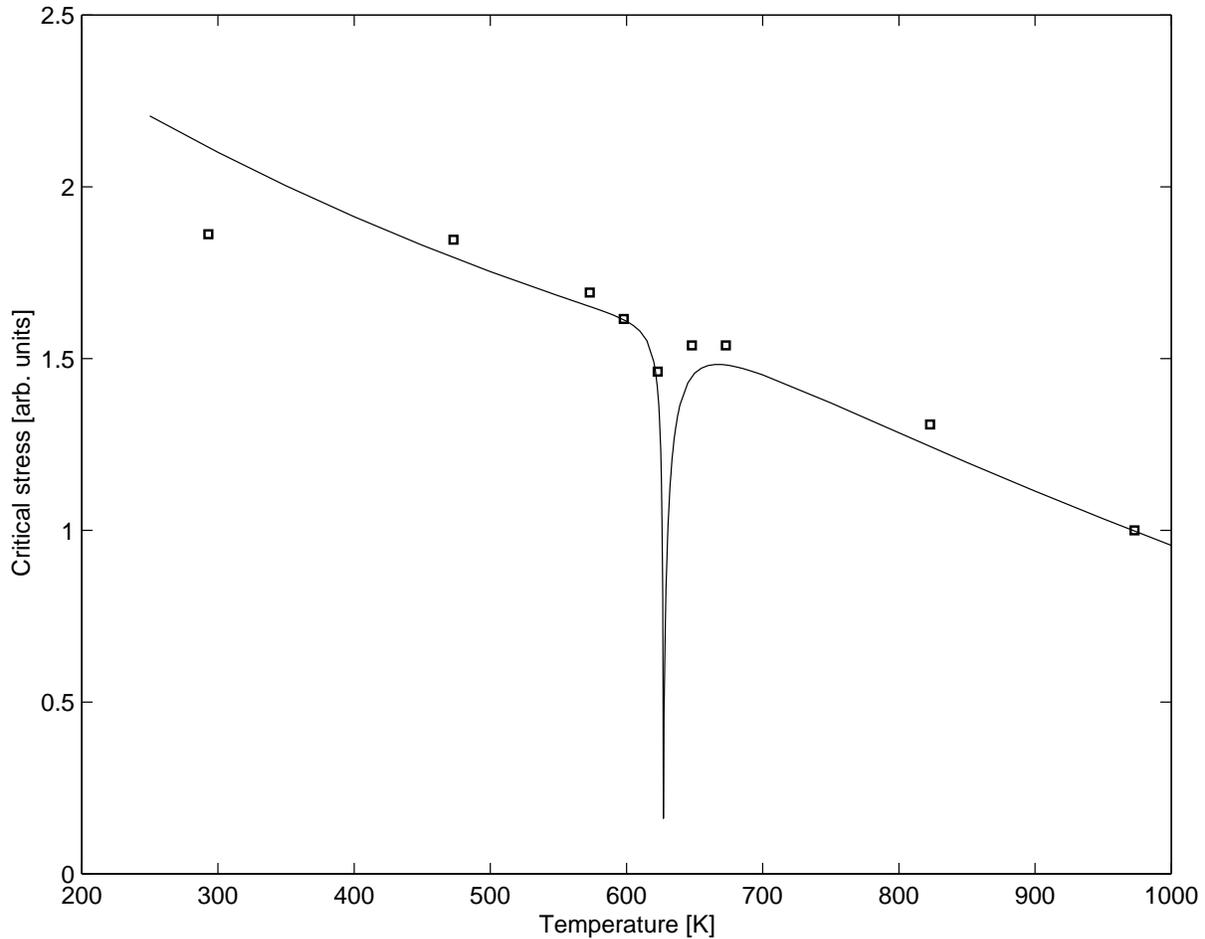} \caption{Temperature
dependence of the critical stress in Ni.}}
\end{figure}
The theoretical curve is plotted using the following parameters
for Ni: $T_C = 627$K, $M_0 = 510$G, $p = 0.606$ (Kittel, 1986),
$\beta = 0.33$, $\gamma = 1.33$ (Kadanoff etal, 1967). The
effective number, p, of Bohr magnetons in Ni is rather small and,
hence, one gets a small Curie constant $C = 7\times 10 ^{-3}K$.
The remaining parameters, $T_0 = 2070$K, $B_0 = 0.49$T, $V^* =
2.64 \times 10^{-22}$cm$^3$, are determined by fitting equation
(\ref{seg2}) to the experimental data by Zackay and Hazlett (1953)
for their samples with 2\% deformation. One observes a reasonable
general agreement between the theory and experiment. However, the
measurements were carried out with a $\sim$ 20K interval between
the points. As a result the sharp minimum in the temperature
dependence of the critical stress near the Curie point, predicted
by our theory, might have been overlooked. A relatively small,
$\sim$ 15\%, decrease of the critical stress in this region is
indicative of a possible existence of a sharper minimum in this
temperature range.

\section{Temperature Dependence of the Microhardness: Gadolinium as
a test case}\label{Ga}

This section discusses the behavior of crystal microhardness near
the Curie point. The results are compared with the available
experimental data on gadolinium. Currently there is no microscopic
theory of microhardness, and hence, no microscopic theory of the
influence of a magnetic field on it. Nevertheless, the dependence
of the microhardness on a magnetic field can be estimated by means
of the following simple considerations. It is known that the
microhardness $H$ varies inversely proportionally to the
plasticity --- the higher the plasticity, the lower the hardness.
On the other hand, the plasticity is directly connected to the
dislocation path length. The magnetic field dependence of $L(B)$
is given by equation (\ref{fpl}). Hence, we assume that the
average microhardness of a crystal as
\begin{equation}\label{mh1}
H = H_0\frac{L_0}{L(B)}
\end{equation}
where $H_0$ is the microhardness in the absence of a magnetic
field.

Assuming an isotropic distribution of the orientations one gets
\begin{equation}\label{mh2}
H =  H_0 \overline{f}(\overline B, \langle \Delta B^2 \rangle)
\end{equation}
where the function $\overline{f}(\overline B, \langle \Delta B^2
\rangle)$ is calculated in Appendix. Similarly to the previous
section, we use equations (\ref{aver2}), (\ref{suscept}), and
(\ref{aver3}) in order to get equation (\ref{mh2}) for the
temperature dependence of the microhardness in the vicinity of the
Curie temperature.

A comparison of this theoretical result with the experimental data
available for the microhardness of gadolinium is presented in
Figure 2.
\begin{figure}[htb]
{\includegraphics[width=16cm]{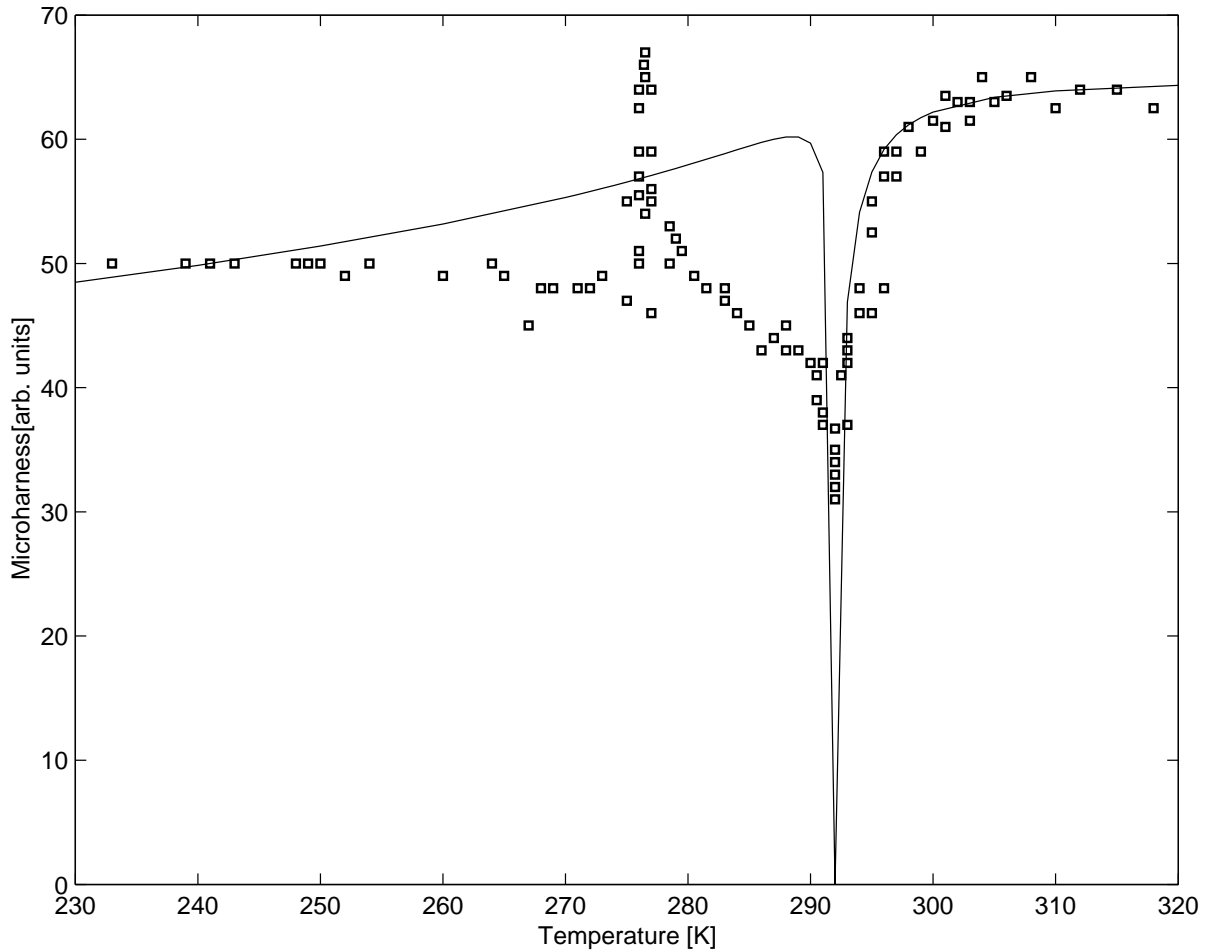
} \caption{Temperature dependence of the microhardness of Gd.}}
\end{figure}
The Curie constant $C$ for gadolinium can be found using equation
(\ref{curie}) and the values of the relevant parameters: $n=3.02
\times 10 ^{22}$cm$^{-3}$, $p = 7.10$, $M_0 = 2010$G (Kittel,
1986), which lead to $C = 0.33$K. The Curie temperature for
gadolinium is $T_C = 292$K (Nabutovskaya, 1969). The critical
indices are $\beta = 0.3265$ and $\gamma = 1.239$ (Aliev etal,
1988).

The value of the parameter $B_0$ could have been possible to
determine from the magnetic field dependence of the plasticity.
Unfortunately, we do not know about such measurements in
gadolinium. However, magneto- (Al'shits etal, 1990) and
electroplastic (Okazaki etal, 1979) effects have been measured in
zinc and titanium, who, similarly to gadolinium, possess hcp
structures. Fitting these results to the theory (Molotskii, 2000)
provides $B_0 = 0.70$T for Zn and $B_0 = 0.94$T for Ti.

The theoretical curve in Figure 2 uses the values $B_0$ and $V^*$
as fitting parameters. The obtained parameter $B_0 = 0.87$T for
gadolinium appears to be closed to its experimentally measured
values for Zn and Ti. The characteristic volume is chosen $V_0 =
1.85\times 10^{-20}$cm$^3$. It corresponds to a sphere with the
radius 17\AA, about four interatomic spacings. The results are not
very sensitive to the variation of this volume, which can be
chosen smaller (up to $2\times 10^{-21}$cm$^3$) without damaging
the overall agreement.

\section{Conclusion}

Accounting for the internal magnetic induction of ferromagnets and
for its fluctuations leads us to the conclusion that various
plastic properties of crystals should be singular near the Curie
point. The critical behavior of the magnetic susceptibility is
transferred to the temperature dependence of the critical stress
and microhardness. It would be interesting to measure other
plasticity characteristics, such as dislocation path length $L$ or
plastic strain rate $\dot{\varepsilon}$ near the Curie point in
order to look for a possible singular behavior, which follows from
the theory proposed in this paper. For example, one may expect
that carrying out measurements similar to those of Zackay and
Hazlett, 1953, on the temperature dependence of the critical
stress in Ni, but with a smaller temperature interval a sharp
minimum at $T=627$K may be observed.

An external magnetic field is known to smear out the phase
transition, to decrease the magnetic susceptibility, and to
suppress fluctuations near $T_C$. As shown by Dan'kov etal (1998)
a 0.5T magnetic field diminishes the magnetic susceptibility near
the Curie point in gadolinium by an order of magnitude. This
should lead to a nearly complete suppression of the magnetization
fluctuations and, hence, to a suppression of the singular
temperature dependence of microhardness. As for nickel, a stronger
field about 3T (Hischler and Rocker, 1966) is necessary to
suppress the fluctuations, hence correspondingly higher field will
be necessary to influence the singular behavior of the critical
stress in nickel near the Curie point.

We may conclude that an experimental observation of a suppression
of the above singularities in the temperature dependencies of
plasticity characteristics in a magnetic field will be an
important experimental evidence in favour of the theoretical
approach to the plasticity of ferromagnets presented in this
paper.

{\bf Acknowledgment}

We are grateful to D. Fleurov for the help with the computation.

\appendix
\section{Averaging for an easy axis cubic crystal}
\label{appenA}

The averaging is carried out assuming a cubic symmetry. It is
certainly relevant to Ni. As for the hexagonal Gd, its anisotropy
is rather weak and the result obtained here, will be also applied
for this crystal. We have to average the quantity
\begin{equation}\label{A.1}
  f(B) = \frac{1}{1 + \displaystyle\frac{{\bf B}^2}{B_0^2}},
\end{equation}
where ${\bf B} = \overline{\bf B} + \Delta{\bf B}$, over the
distribution (\ref{fluct1}). Disregarding a possible anisotropy of
the fluctuations of the magnetic induction in the vicinity of the
dislocation -- obstacle bonds this averaging can be presented as
$$\langle f(B) \rangle \equiv \overline{f}(\overline B, \langle
\Delta B^2 \rangle) = \frac{1}{(2\pi)^{1/2}V^*\langle\Delta B^2
\rangle^{3/2}} \int_{V^*}^\infty dV \int_0^\infty \Delta
B^2d\Delta B \int_{-1}^1 \frac{d\cos\theta}{2}\times$$
\begin{equation}\label{A.2}
 \frac{1}{1 +
\displaystyle\frac{1}{B_0^2} (\overline B~^2 + \Delta B^2 + 2
\overline{B} \Delta B \cos\theta)}\exp\left\{-\frac{\Delta B^2 V
}{2\langle\Delta B^2 \rangle V^*}\right\}
\end{equation}

Now integration over $\Delta B$ is carried out by parts and the
averaging (\ref{A.2}) is represented as
$$\overline{f}(\overline B, \langle \Delta B^2 \rangle) =$$
$$
\frac{1}{2(2\pi)^{1/2}\langle\Delta B^2 \rangle^{1/2}}
\int_{V^*}^\infty dV \frac{1}{V}\int_{-\infty}^\infty d\Delta B
\frac{1}{1 + \displaystyle\frac{1}{B_0^2} (\overline B + \Delta
B)^2} \exp\left\{-\frac{\Delta B^2 V}{2\langle\Delta B^2 \rangle
V^*}\right\} +
$$
\begin{equation}
\label{A.3}\frac{1}{2(2\pi)^{1/2} \langle\Delta B^2 \rangle^{1/2}
\overline{B}} \int_{V^*}^\infty \frac{dV}{V} \int_{-\infty}^\infty
d\Delta B \frac{\Delta B}{1 + \displaystyle\frac{1}{B_0^2}
(\overline B + \Delta B)^2} \exp\left\{-\frac{\Delta B^2
V}{2\langle\Delta B^2 \rangle V^*}\right\}
\end{equation}
or
\begin{equation}\label{A.4}
\overline{f}(\overline B, \langle \Delta B^2 \rangle) =
\frac{1}{2\sqrt{2\pi}} \int_{1}^\infty
\frac{dv}{v}\int_{-\infty}^\infty dx \frac{a^2}{a^2 + (b + x)^2}
(1 + \frac{x}{b}) \exp\left\{-\frac{x^2 v}{2}\right\}
\end{equation}
where the notations $a=B_0/\sqrt{\langle\Delta B^2\rangle}$, $b =
\overline{B}/\sqrt{\langle\Delta B^2\rangle}$, $v = V/V^*$, and $x
= \Delta B/\sqrt{\langle\Delta B^2\rangle}$ are introduced. The
identity
\begin{equation}\label{A.5}
\frac{1}{a \pm i (b + x)} = \int_0^\infty dp \exp\{-p[a \pm i(b +
x)]\}
\end{equation}
is substituted into (\ref{A.4}). Now integration over $x$ is
carried out and a new integration variable $z = p/\sqrt{v}$ is
introduced instead of $v$. After that, $p$ is substituted for $pa$
and one gets
$$\overline{f}(\overline B, \langle \Delta B^2 \rangle) =
\frac{B_0}{\sqrt{\langle\Delta B^2\rangle}} \int\limits_{0}^\infty
dp \times$$
\begin{equation}\label{A.6}
 \left\{\sqrt{\frac{\pi}{2p}} e^{-p}
\Phi\left(\frac{\sqrt{\langle\Delta
B^2\rangle}p}{B_0\sqrt{2}}\right) \left[
\cos\left(\frac{\overline{B}p}{B_0}\right) -
\frac{B_0}{4\overline{B}p}
\sin\left(\frac{\overline{B}p}{B_0}\right)\right] +
\frac{B_0}{4\overline{B}p} e^{-p
 - \displaystyle \frac{\langle \Delta B^2 \rangle p^2}{2
B_0^2}}\sin\left(\frac{\overline{B}p}{B_0}\right)\right\}
\end{equation}
where
\begin{equation}\label{A.7}
  \Phi(x) = \frac{2}{\sqrt{\pi}}\int_0^x e^{-t^2}dt
\end{equation}
is the probability integral.

The value of $\overline{f}(\overline B, \langle \Delta B^2
\rangle)$ can be analytically estimated in the limit of small
fluctuations, when $\sqrt{\langle\Delta B^2\rangle} \ll B_0 $. The
integral (\ref{A.6}) converges for small values of $p$ and the
expansion $\Phi(x)\approx 2x/ \sqrt{\pi}$ can be used. It
converges to a rather obvious result
\begin{equation}\label{A.8}
\overline{f}(\overline B, \langle \Delta B^2 \rangle) =
\frac{B_0^2}{B_0^2 + \overline{B}^2}.
\end{equation}

For large fluctuations when $\sqrt{\langle\Delta B^2\rangle} \gg
B_0 $ the integration (\ref{A.6}) includes a broad range of large
values of $p$. The probability integral $\Phi(x)$ rapidly
converges to one for $x \gg 1$. Then we may exclude the range $p <
p_0 =\displaystyle \frac{B_0\sqrt{2}}{\sqrt{\langle \Delta B^2
\rangle}}$ whose contribution to the integral is of the order of
$\displaystyle \left(\frac{ B_0}{\sqrt{\langle\Delta B^2\rangle}}
\right)^2$ and integrate for $p > p_0$. Then one gets
$$\overline{f}(\overline B, \langle \Delta B^2 \rangle) = -
\frac{B_0}{2\sqrt{\langle\Delta B^2\rangle}}
\sqrt{\frac{\pi}{2}}\left[\mbox{Ei}(-\sqrt{2}\frac{B_0 +
i\overline{B}}{\sqrt{\langle\Delta B^2\rangle}}) +
\mbox{Ei}(-\sqrt{2}\frac{B_0 - i\overline{B}}{\sqrt{\langle\Delta
B^2\rangle}}) \right] \approx$$
\begin{equation}\label{A.9}
\sqrt{\frac{\pi}{2}} \frac{B_0}{\sqrt{\langle\Delta B^2\rangle}}
\left[\ln(\frac{ \langle\Delta B^2\rangle}{B_0^2 + \overline{ B }
~ ^2}) - C - \frac{1}{2} \ln 2\right]
\end{equation}
where $C =0.577$ is the Euler number. Ei($x$) is the integral
exponential function.\\

\centerline{\bf REFERENCES}

\noindent Aliev, Kh.K., Kamilov, I.K., and Omarov, A.M.,
    Zh.Eksp.Teor.Fiz., {\bf 94}, 153 (1988) [Sov.Phys.JETP, {\bf 67},
    2262 (1988)].

\noindent Al'shits, V.I., Darinskaya, E.V., Getkina, I.V., and
    Lavrentiev, F.F., Kristallografiya, {\bf 35}, 1014 (1990)
    [Sov.Phys.Crystallogr., {\bf 35}, 597 (1979)]

\noindent Bolling, C.F., and Richman, R.H., Philos.Mag., {\bf 19},
    247, (1969).

\noindent Dan'kov, S.Yu., Tishin, A.M., Pecharsky, V.K., and
    Gschneider, K.A., Jr., Phys.Rev. B {\bf 57}, 3478 (1998)

\noindent Echigoya, J., Hayashi, S., and Yamamoto, Y.,
    Phys.Stat.Sol., (a), {\bf 14}, 463 (1973).

\noindent Echigoya, J., and Hayashi, S., Phys.Stat.Sol., (a), {\bf
    55}, 279 (1979).

\noindent Flor, H., Gudladt, H.J., and Schwink, Ch., Acta Metal.,
    {\bf 28}, 1611 (1980).

\noindent Friedel, J.,{\em Dislocations}, (Pergamon, Oxford,1964).

\noindent Gulayev, A., and Svistunova, E.L., Scr.Mater., {\bf 35},
    501 (1996).

\noindent Haasen, P., in {\em Physical Metallurgy}, edited by R.W.
    Cahn and P.Haasen (North-Holland, Amsterdam, 1983), Part II, pp.
    1341-1409.

\noindent Hischler, W., and Rocker, W., Z.Angew.Phys., {\bf 21},
    386 (1966)

\noindent Kadanoff, L.P., G\"otze, W., Hamblen, D., Hecht, R.,
    Lewis, E.A.S., Palciauskas, V.V., Rayl, M., Swift, J., Aspens, D.,
    and Kane, J., Rev.Mod.Phys., {\bf 39}, 395 (1967)

\noindent Kittel, C., {\em Introduction to Solid State Physics},
    6-th edition (J.Wiley \& Sons, New York) 1986.

\noindent Landau, L.D., and Lifshitz, E.M., {\em Statistical
    Physics}, 3rd Edition, Part 1 (Pergamon Press, Oxford, 1980),
    Ch.14.

\noindent Ma, Shang-Keng, {\em Modern Theory of Critical
    Phenomena}, (Benjamin, London, 1976)

\noindent Maksimova, G.A., and Maiboroda, V.P., Metal Science and
Heat Treatment, {\bf 34}, 339 (1992).

\noindent Molotskii, M., Mater.Sci.Eng., A{\bf 287}, 248 (2000).

\noindent Molotskii, M., and Fleurov, V., Phys.Rev.Lett., {\bf
    78}, 2779 (1997).

\noindent Molotskii, M., and Fleurov, V., Phys.Rev., B {\bf 63},
    184421 (2001)

\noindent Nabutovskaya, O.A., Fiz.Tverd.Tela (Leningrad), {\bf
    11}, 1434 (1969) [Sov.Phys. Solid State, {\bf 11}, 1172 (1969)].

\noindent Nabutovskaya, O.A., Fiz.Tverd.Tela (Leningrad), {\bf
    13}, 3130 (1971) [Sov.Phys. Solid State, {\bf 13}, 3130 (1971)].

\noindent Okazaki, K., Kagawa, M., and Conrad, H., Scr.Metal.,
    {\bf 13}, 473 (1979)

\noindent Retat, I., Steffens, Th., and Schwink, Ch.,
    Phys.Stat.Sol;, (a), {\bf 92}, 507 (1985).

\noindent Retat, I., Phys.Stat.Sol., (a), {\bf 99}, 121 (1981).

\noindent Seeger, A., Kronm\"uller, H., Riegre, H., and Tr\"auble,
    H., J.Appl.Phys., {\bf 35}, 740 (1964)

\noindent Spichkin, Y.I., Tishin, A.M., and Gschneider, K.A., Jr.,
    J.Magn.Magn.Mater., {\bf 204}, 5 (1999).

\noindent Vonsovskii, S.V., {\em Magnetism}, (J.Wiley \& Sons, New
    York, 1974), Vol. 2.

\noindent Wolfenden, A., Z.Metallkunde, {\bf 69}, 308 (1978).

\noindent Zackay, V.F., and Hazlett, T.H., Acta Metal., {\bf 1},
    624 (1953).

\end{document}